\documentclass[aps,prl,superscriptaddress,twocolumn,reprint]{revtex4-1}
\usepackage{bm, amsmath, amsthm, graphicx,graphics, times, color, amssymb, epstopdf, txfonts, subfigure, tabularx, float}
\newcommand{\be}{\begin{equation}}
\newcommand{\ee}{\end{equation}}
\newcommand{\bea}{\begin{eqnarray}}
\newcommand{\eea}{\end{eqnarray}}

\begin{document}

\title{Entanglement structure of the two-channel Kondo model}

\author{Bedoor Alkurtass}
\affiliation{Department of Physics and Astronomy, University College London, Gower Street, London WC1E 6BT, UK}
\affiliation{Department of Physics and Astronomy, King Saud University, Riyadh 11451, Saudi Arabia}

\author{Abolfazl Bayat}
\affiliation{Department of Physics and Astronomy, University College London, Gower Street, London WC1E 6BT, UK}

\author{Ian Affleck}
\affiliation{Department of Physics and Astronomy, University of British Columbia, Vancouver, British Columbia V6T 1Z1, Canada}

\author{Sougato Bose}
\affiliation{Department of Physics and Astronomy, University College London, Gower Street, London WC1E 6BT, UK}

\author{Henrik~Johannesson}
\affiliation{Department of Physics, University of Gothenburg, SE 412 96 Gothenburg, Sweden}

\author{Pasquale Sodano}
\affiliation{International Institute of Physics, Universidade Federal do Rio Grande do Norte, 59078-400 Natal-RN, Brazil}
\affiliation{Departemento de Fis\'ica Teorica e Experimental, Universidade Federal do Rio Grande do Norte, 59072-970 Natal-RN, Brazil}

\author{Erik S. S{\o}rensen}
\affiliation{Department of Physics and Astronomy, McMaster University, Hamilton, Ontario L8S 4M1, Canada}

\author{Karyn Le Hur}
\affiliation{Centre de Physique Th\'{e}orique, Ecole Polytechnique, CNRS, Universit\'{e} Paris-Saclay, 91128 Palaiseau Cedex, France}

\begin{abstract}
Two electronic channels competing to screen a single impurity spin, as in the
two-channel Kondo  model, are expected to generate a ground state with
nontrivial entanglement structure. We exploit a spin-chain representation of
the two-channel Kondo model to probe the ground-state block entropy,
negativity, tangle, and Schmidt gap, using a density matrix renormalization
group approach.
In the presence of symmetric coupling to the two channels we
confirm field-theory predictions for
the boundary entropy difference, $\ln (g_{UV}/g_{IR})=\ln(2)/2$, between
the ultraviolet and infrared  limits and the
leading $\ln(x)/x$ impurity correction to the block entropy.
The impurity entanglement, $S_{\text{imp}}$, is shown to scale with the
characteristic length $\xi_{2CK}$.
We show that both  the Schmidt gap and the entanglement of the impurity with one of the channels $-$ as measured by
the negativity$-$ faithfully serve as order parameters for the
impurity quantum phase transition appearing as a function of channel asymmetry,
allowing for explicit determination of critical exponents, $\nu\!\approx\! 2$ and
$\beta \!\approx\! 0.2$. Remarkably,  we
find the emergence of tripartite entanglement only in
the vicinity of the critical channel-symmetric point.
\end{abstract}


\maketitle


{\em Introduction.-} The Kondo effect is one of the most intriguing effects in
quantum many-body physics. At low temperatures, a localized magnetic impurity
is screened by the conduction electrons leading to the formation of many-body
entanglement. A generalization of the Kondo model was introduced by Nozi\`eres
and Blandin \cite{nozieres1980kondo}, where another channel of electrons is
also coupled to the impurity. This
is the well-known two-channel Kondo (2CK) model, for which various results were obtained using  Bethe ansatz
\cite{andrei1984solution,tsvelick1984solution,tsvelick1985entropy},  conformal field theory
\cite{affleck1991critical,affleck1993exact} (CFT), bosonization
\cite{emery1992mapping,GiamarchiShraimanClarke,SenguptaGeorges} and entanglement of formation \cite{sim-PRL-2015}. This model is
very different from the one-channel Kondo (1CK) model as the two
channels compete to screen the spin-$1/2$ impurity, leading to an
``overscreened'' residual spin interacting with the electrons
\cite{affleck1991critical}. This leads to non-trivial properties including a
residual zero-temperature impurity entropy and a logarithmic behavior of
magnetic susceptibility and specific heat. However, channel symmetry is
crucial; even the smallest asymmetry leads to screening of the impurity by the
channel with the stronger coupling \cite{nozieres1980kondo},
and as the channel asymmetry is varied, an impurity quantum phase transition (IQPT) occurs
at the symmetric point, corresponding to the 2CK model.

Intensive research has been carried out to investigate the
thermodynamics and the transport properties of the
2CK model \cite{nozieres1980kondo,andrei1984solution,tsvelick1984solution,affleck1991critical,affleck1993exact,emery1992mapping,GiamarchiShraimanClarke,SenguptaGeorges,eggert1992magnetic,affleck1994,eggert2001phase,cragg1980,sacramento1991low,affleck1992,gan1993,andrei1995,fabrizio1995,zarand2000,yotsuhashi2002,toth2008,mitchell2012universal,mitchell2011real,RC2CK,eggert1992magnetic,affleck1994,eggert2001phase}.
Experimentally, signatures of the 2CK model have been observed in
mesoscopic structures \cite{Potok,Gleb,Keller,FPierre}.
Still, the real-space entanglement
structure and the imprints of the two
distinct length scales
$\xi_{\text{2CK}}\sim u/T_{\text{2CK}} $ and $\xi^{\ast}\sim u/T^{\ast}$
with $u$ the spin velocity-- implied by the
known crossover energy scales $T_{\text{2CK}}$ (2CK temperature)
and $T^{\ast}$ (critical crossover in the channel-asymmetric case)
\cite{affleck1991critical,mitchell2011real} $-$ have not yet been unraveled.
A way forward
is to use
%
a spin-chain representation of the 2CK model~\cite{eggert1992magnetic,affleck1994},
which allows for efficient Density Matrix Renormalization Group (DMRG)
computations~\cite{sorensen20071CK,sorensen2007QIE,affleck2009Review,bayat2010negativity,bayat2012entanglement}
($m=100-1024$ states kept)
to uncover the ground state entanglement properties.

In this letter, we show how the implementation of this scheme  allows for a detailed study of the entanglement in the
2CK ground state and the  IQPT between the two channel-asymmetric 1CK phases.
Specifically, we present results for
the impurity entanglement
entropy  \cite{sorensen20071CK,sorensen2007QIE}, the negativity
\cite{Kim-Negativity-2000,vidal2002}, the Schmidt gap
\cite{de2012entanglement,bayat2014order}, and the tripartite entanglement \cite{tangle-wootters,tangle-Fan}.  At the channel-symmetric 2CK point
we show that $\xi_{\text{2CK}}$ can be interpreted as a  dynamically generated cutoff length
by demonstrating scaling of the impurity entanglement entropy.
A detailed analysis allows us to extract the
two-channel boundary entropy difference $\ln (g_{UV}/g_{IR})=\ln(2)/2$, between the ultraviolet and infrared limits~\cite{AffleckLudwig1991},
 as well as the leading correction $\ln(x)/x$, for block sizes $x \gg \xi_{\text{2CK}}$
\cite{eriksson2011}.
In addition, we show that the negativity and the Schmidt gap act as order parameters for the IQPT,
enabling us to predict, via finite-size scaling, the pertinent critical exponents.
Finally, we compute the tangle
\cite{tangle-wootters,tangle-Fan} and show that tripartite entanglement emerges {\em only} in the vicinity of
the critical point.

\begin{figure}[t]
\centering
\includegraphics[width=.42\textwidth] {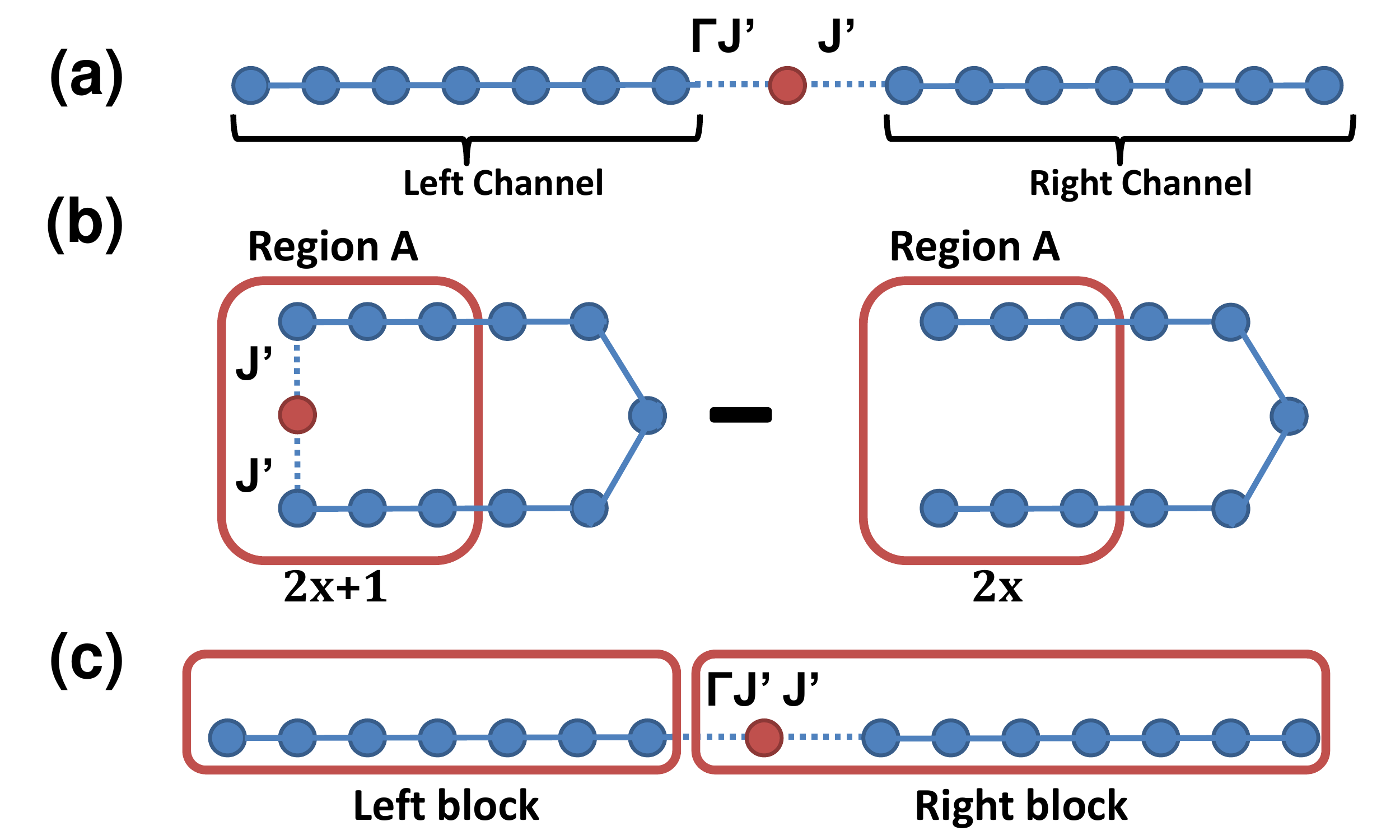}
   \caption{{
     \protect\footnotesize
       (Color online)
     (a) Kondo spin chain with  a spin-1/2 impurity coupled to its left and right channels by $\Gamma J'$ and $J'$, respectively.
       For $\Gamma=1$ the impurity is screened by both channels representing the 2CK model while for $\Gamma\neq 1$ 1CK physics emerges.
   (b) The impurity entropy $S_{\text{imp}}$ is computed as the difference between the entropy of region $A$ with and without the impurity.
   (c) Partitioning of the system for computing the Schmidt gap.}}
\label{SCKM2CK}
\end{figure}

{\em Spin-chain representation.-} We consider two open Heisenberg chains coupled to a single spin$-1/2$ impurity as shown in Fig.~\ref{SCKM2CK}(a). The open chain Hamiltonian is given by
\begin{eqnarray} \label{eq:Hamiltonian}
\nonumber H_{OBC}&=&\sum_{m=L,R}\Bigg[J'_{m}\left(J_1\boldsymbol{\sigma}^0\cdot\boldsymbol{\sigma}_{m}^{1}+J_2\boldsymbol{\sigma}^0\cdot\boldsymbol{\sigma}_{m}^{2}\right)\\
&+ &J_1\sum_{l=1}^{N_{m}-1}\boldsymbol{\sigma}_{m}^l\cdot\boldsymbol{\sigma}_{m}^{l+1}+J_2 \sum_{l=1}^{N_{m}-2}\boldsymbol{\sigma}_{m}^l\cdot\boldsymbol{\sigma}_{m}^{l           +2}\Bigg]\, ,
\end{eqnarray}
where $\boldsymbol{\sigma}^0$ and $\boldsymbol{\sigma}^l_{m}$ represent the vector of Pauli matrices for the impurity spin and the spin at site $l$ in channel $m$, respectively, and $N_m$ is the number of spins in
 chain $m$ making the total number of spins $N=N_L+N_R+1$. We choose the nearest-neighbor coupling $J_1$ to be unity and the next-nearest-neighbor coupling $J_2=J_2^c$ (with $J_2^c=0.2412 J_1$) so as to remove marginal coupling effects \cite{laflorencie2008kondo,bayat2012entanglement}. In this work, we set the Kondo coupling $J'_L = \Gamma J'$ and $J'_R=J'$, with $\Gamma=J'_L/J'_R$ keeping $J'_m<1$.
The Hamiltonian (\ref{eq:Hamiltonian}) has been
introduced in Ref.~\cite{eggert1992magnetic,affleck1994} as a representation of the spin
sector of the 2CK model when $\Gamma=1$. For  further justifications, see the Supplemental Material. For any $\Gamma\neq 1$
1CK physics emerges. For the case of $\Gamma= 1$, we also use a periodic chain, as shown on the left side of the Fig.~\ref{SCKM2CK}(b), by adding the following terms:
\begin{eqnarray} \label{eq:Hamiltonian_PBC}
H_{PBC}= H_{OBC}&+&J_1 \boldsymbol{\sigma}_L^{N_L}\cdot \boldsymbol{\sigma}_R^{N_R}
+J'J_2 \boldsymbol{\sigma}_L^1\cdot \boldsymbol{\sigma}_R^1\nonumber\\
&+&J_2\left( \boldsymbol{\sigma}_L^{N_L}\cdot \boldsymbol{\sigma}_{R}^{N_R-1}+
 \boldsymbol{\sigma}_L^{N_L-1}\cdot \boldsymbol{\sigma}_R^{N_R}\right).
\end{eqnarray}
Again $N=N_L+N_R+1$, and at $J'=1$ we obtain a uniform periodic chain which presents significant
advantages \cite{SupMat}.
In the limit of $N\rightarrow \infty$ the two boundary conditions are equivalent.
For $H_{OBC}$ the parity of $N_L = N_R$  is crucial~\cite{toappear} but here we only study $N_L=N_R$ {\it odd}, however,
for $H_{PBC}$ it is the parity of $N$ that matters~\cite{toappear} and we only study $N$ even ($N_L= N_R\pm 1$) which makes
the parity effects compatible for $H_{OBC}$ and $H_{PBC}$.

{\em Impurity entanglement entropy.-}
We first study the channel-symmetric case,
$\Gamma=1$, with $\xi_{\text{2CK}}$ being the only relevant length
scale in the problem. We consider the von Neumann entropy, $S_A(J',x,N)=-\text{Tr}\rho_A\log\rho_A$ with $\rho_A$ the reduced density matrix of a
region $A$ which includes the impurity spin
and $x$ spins on either side of it. $N$ is the total number of spins in
the system, including the impurity. We consider an even {\it periodic} system, using $H_{PBC}$ as shown in Fig.~\ref{SCKM2CK}(b). This boundary condition should
not affect our results as long as $x\ll N/2$ \cite{SupMat}.
 Similar to the single-channel case
\cite{sorensen20071CK,sorensen2007QIE} the entanglement entropy behaves very
differently in the two limits  $x\ll \xi_{\text{2CK}}$ and $x\gg
\xi_{\text{2CK}}$, with $\xi_{\text{2CK}}\sim e^{a/J'}$ growing exponentially
as $J'\to 0$ (for some constant $a$).
In what follows we shall show how to pinpoint the impurity contribution, $S_{imp}$, to the von Neumann entropy. By doing so, we provide a direct "quantum probe" of the boundary entropy predicted by CFT \cite{AffleckLudwig1991}, with no reference to the thermodynamic entropy.

Let's first consider the $N\to \infty$ limit.
When $J'=1$, we simply have a uniform periodic chain with region $A$ consisting of $2x+1$ sites. Then, using the fact that the central charge $c=1$, the
entanglement entropy for region $A$ of a periodic chain is predicted to be, from CFT \cite{calabrese2004}
\begin{equation}\label{Sper}
S_A(J'=1,x,N)={1\over 3}\ln (2x+1)+s_1
\end{equation}
for a non-universal constant $s_1$.  For {\it finite} but large $N$ {\it even},
   we expect the limit of $J'\to 0^+$, $x\ll N$  (which is different from the case where the impurity is absent) to give:
\begin{equation}\label{Ln2_addition}
S_A(J'\rightarrow 0^+,x,N)=S_A(x,N-1)+\ln 2
\end{equation}
where $S_A(x,N-1)$ represents the entropy of region $A$ when the impurity
is absent but the region still consists of $x$ spins from each channel (so the total length is $N-1$) as shown on the right side of Fig.~\ref{SCKM2CK}(b). The
additional $\ln 2$ entanglement entropy in Eq.~(\ref{Ln2_addition}) is the
impurity contribution and can be understood by observing that a spin chain with
an even number of sites has a spin zero ground state for any  $J'>0$ no matter
how small. In a valence bond picture of the $N$ even ground state there will
always be an (impurity) valence bond (IVB) connecting the impurity spin to another spin in the
system, although the IVB becomes very long in the small $J'$ limit \cite{sorensen20071CK,sorensen2007QIE}.
Intuitively, this long IVB adds an extra $\ln 2$ to  $S_A(J'\to 0^+,x,N)$.
The interesting case of $N$ odd will be considered elsewhere~\cite{toappear}.

In the absence of an impurity, as long as $x\ll N/2$,  the entropy of region $A$ is the sum of the entropy of two equal blocks at either end
of an open chain as shown in the right part of Fig.~\ref{SCKM2CK}(b). In this case the open boundaries
induce an alternating term in the entanglement entropy~\cite{laflorencie2006} and we therefore only focus on the uniform part, $S^u$ finding~\cite{calabrese2004,Zhou2006}
\begin{equation}\label{SA_Jimp-0}
 S^u_A(x,N-1) = 2 [ {1\over 6}\ln (2x)+\frac{s_1}{2}+\ln g ],
\end{equation}
where $s_1$ is the same non-universal constant appearing in Eq.~(\ref{Sper}) and $\ln g$ is a universal
term arising from a non-integer ``ground-state degeneracy'', $g$~\cite{AffleckLudwig1991}.

The difference between the two entropies of the two extreme regimes will be
\begin{equation}\label{Difference_entropy}
 S_A(J'=1,x,N)-S^u_A(J'\to 0^+,x,N)=-2\ln g-\ln 2+O(1/x).
\end{equation}
Using the mapping of the spin-chain system onto the 2CK model, we associate $J'\to 0^+$ with the weak coupling ultraviolet fixed point and $J'\to 1$
with the infrared fixed point.  Hence we expect
\begin{equation}
S_A(J'=1,x,N)-S^u_A(J'\rightarrow 0^+,x,N)=\ln g_{IR}-\ln g_{UV},
  \label{eq:UVIR}
\end{equation}
where $\ln g_{UV}$ and $\ln g_{IR}$ are the boundary entropies for the
ultraviolet and infrared fixed points. Hence, it follows that the degeneracies of the 2CK model and the open chain {\it must be related} as
$g_{UV}/g_{IR}=2g^2$.
While $g_{UV}=2$, corresponding to the decoupled impurity spin, $g_{IR}$ has
the non-trivial value of $\sqrt{2}$.
On the other hand,  $g$
was predicted, using field theory arguments for the open spin chain to have the
value $2^{-1/4}$ \cite{eggert1992magnetic,affleck1994,oshikawa2007} validating the relation $g_{UV}/g_{IR}=2g^2=\sqrt{2}$.
This constitutes a highly non-trivial check of the spin-chain representation of the 2CK model. We confirm
the result $g=2^{-1/4}$ by extracting $s_1$ from DMRG results for the entanglement
entropy for an even periodic chain, finding $s_1=0.743743$.
We then fit $S_A^u$ for a single open chain of length $N$ to the finite $N$ generalisation of
Eq.~(\ref{SA_Jimp-0})
\begin{eqnarray} \label{SO}
S^u_A(x,N)&=&2\left[ {1\over 6}\ln  [(2N/\pi )\sin (\pi x/N)]+{s_1\over 2}+\ln g \right]  \nonumber \\
&+&  {\alpha\over N}[2+\pi (1-2x/N)\cot (\pi x/N)].
\end{eqnarray}
Here the last term is a correction, behaving as $\alpha/x$ in the $N\to \infty$ limit, calculated in \cite{sorensen20071CK,sorensen2007QIE,toappear} where $\alpha$ is a non-universal parameter.
$S^u_A$ is extracted using a 7-point formula~\cite{toappear,sorensen20071CK,sorensen2007QIE}. With $s_1$ known, this then determines $ \ln g=-0.17328,$ in excellent agreement with $\ln (2^{-\frac{1}{4}})=-0.1732867\ldots$.

\begin{figure}[t]
\centering{
\includegraphics[width=.45\textwidth]{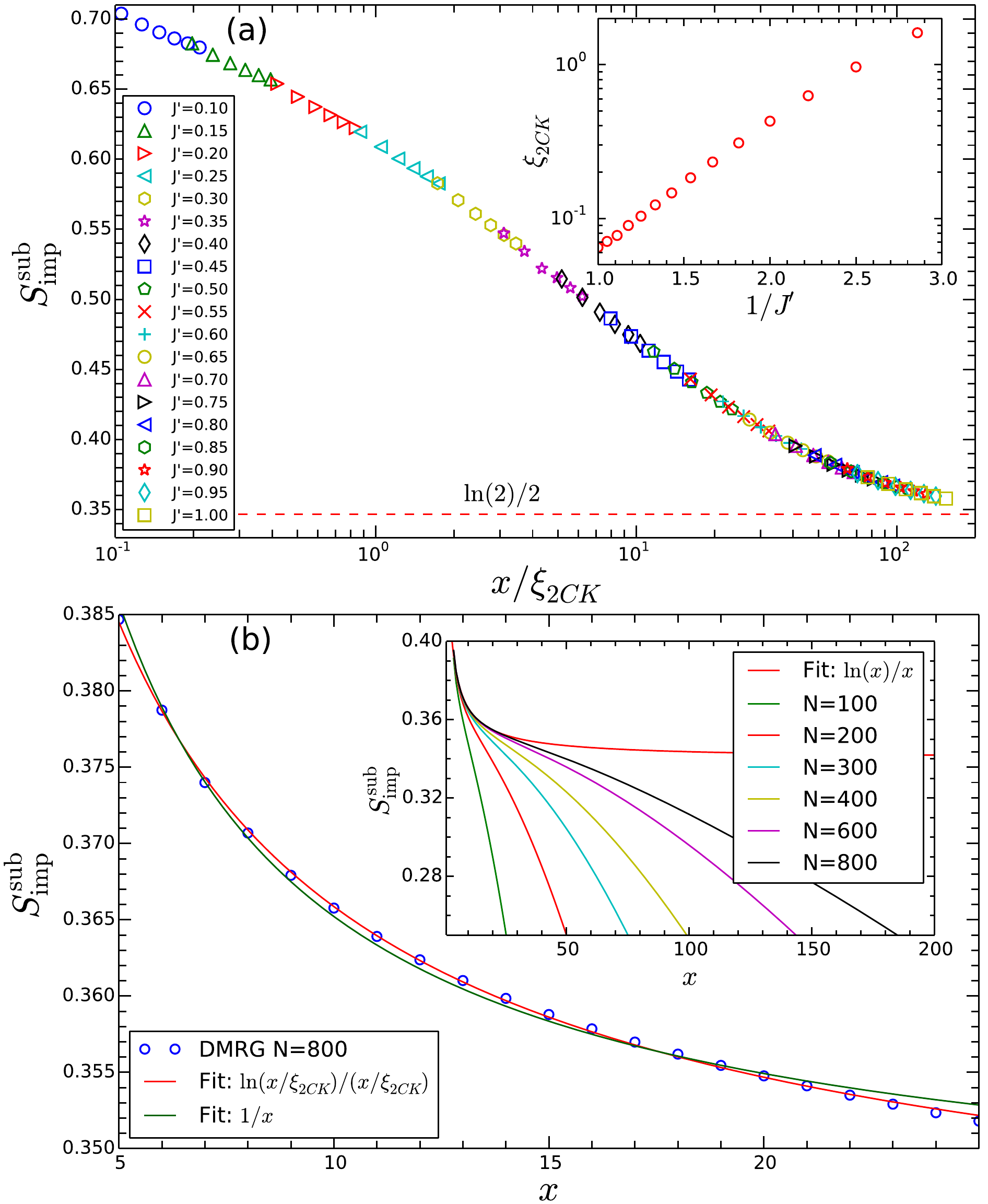}}
 \caption{{
   \protect\footnotesize
     (Color online) (a)
  Scaling of $S_{\rm imp}^{\rm sub}(x/\xi_{\text{2CK}}(J'))$ for fixed $x/N=1/10$ ($N$ even). At $J'=0.9$, $\xi_{\text{2CK}}(J')$ is arbitrarily fixed at $0.07747$ to coincide with the estimate from panel b. Inset: $\xi_{\text{2CK}}(J')$ as a function of $1/J'$.
 (b)  DMRG results for $S_{\rm imp}^{\rm sub}(x;J'=0.9,N=800)$. For $\xi_{2CK}(J')\ll x\ll N/2$
  $S_{\rm imp}^{\rm sub}$ can be fit to the form $A\ln(x/\xi_{2CK})/(x/\xi_{2CK})+B$ (red line) with
  $\xi_{2CK}(J'=0.9)=0.07747$, $A=0.69$ and $B=0.34\sim\ln(2)/2$   significantly better than to
  $\sim 1/x$ (green line). Inset: Convergence to the limiting form at $x\ll N/2$ with $N$. }}
\label{Simp}
\end{figure}

We now show that $\ln(g_{UV}/g_{IR})$ enters as part of the limiting behavior
of the impurity entanglement entropy allowing us to numerically estimate this
boundary entropy difference.  We begin by considering the behaviour of $S_A$
for intermediate values of $J'$.  Most notably, an alternating term appear in
$S_A$ for {\it any} $J'\neq 1$\cite{toappear}. Hence,
by subtracting off the entropy with the impurity
absent~\cite{sorensen20071CK,sorensen2007QIE}, as shown in Fig.~\ref{SCKM2CK}(b), we define the impurity
entanglement entropy using the {\it uniform part} as
\begin{equation}\label{Simp_def}
 S_{\text{imp}}(J',x,N)=S_A^u(J',x,N)- S_A^u(x,N-1).
\end{equation}
The hallmark feature of the characteristic length $\xi_{\text{2CK}}\sim
u/T_{\text{2CK}}$ is that $S_{\text{imp}}$ is a {\it universal} scaling
function of the two variables $x/N$ and $x/\xi_{\text{2CK}}$.  Again, the
parity of $N$ also plays a crucial role~\cite{toappear} but here we only focus
on $N$ {\it even}.  If we fix $x/N=1/10$, $S_{\text{imp}}$ should then be  a
function of the single variable $x/\xi_{\text{2CK}}$.  However, as evident from
Eq. (\ref{SO}) the term proportional to $\alpha$ in $S_A^u(x,N-1)$ gives rise
to corrections to scaling disappearing as $N\to \infty$ with $x/N$ and
$x/\xi_{2CK}$ held fixed.  For clarity, we subtract these corrections from
$S_{\text{imp}}$  obtaining $S_{\rm imp}^{\rm sub}$.  In Fig.~\ref{Simp}(a) we
demonstrate the scaling by collapsing data for many values of $J'$ and $N$ with
fixed $x/N$ onto a single curve by appropriately selecting
$\xi_{\text{2CK}}(J')$.
The expected $\xi_{\text{2CK}}\sim e^{a/J'}$ behavior is also
confirmed (inset of Fig.~\ref{Simp}(a)).
We see that an excellent data collapse appears for a
range of $J'$ and the data approaches fairly closely to
$\ln(2)/2=0.3465$ at large $x/\xi_{\text 2CK}$ .
This limit corresponds to
$J'\to 1$ and
using Eq.~(\ref{Ln2_addition}) and (\ref{eq:UVIR}) we have
$S_{\text{imp}}(J'\to 1)=\ln(2)-\ln(g_{UV}/g_{IR})=\ln(2)/2$ so we can conclude\cite{SupMat}:
\begin{equation}
\ln(g_{UV}/g_{IR})\simeq \ln(2)/2, \ \ g_{UV}/g_{IR}\simeq \sqrt{2}
\end{equation}
providing a firm confirmation of the CFT predictions.

For $x \gg \xi_{\text{2CK}}$ at $N\to \infty$
we are close to the infrared fixed point.  The leading irrelevant operator has dimension 3/2 \cite{affleck1991critical} and is expected to
lead to corrections to the leading $\ln( 2)/2$ behavior of $S_{\text{imp}}$ that in second-order perturbation theory are of the form
 $\delta S_{\text{imp}}\propto \ln( x)/x$ \cite{eriksson2011}, valid in the regime $\xi_{\text{2CK}}\ll x\ll N/2$. Numerically we can confirm this by studying $S_{\text{imp}}^{\text{sub}}$ for $J'\sim 1$
where $\xi_{\text{2CK}}$ is small. This is shown
in Fig.~\ref{Simp}(b)
for $J'=0.9$
where a fit to $\ln(x)/x$ correction is statistically superior to a simpler $1/x$ form over a significant range of $x$.

\begin{figure}[t]
\centering{
\includegraphics[width=0.5\textwidth]{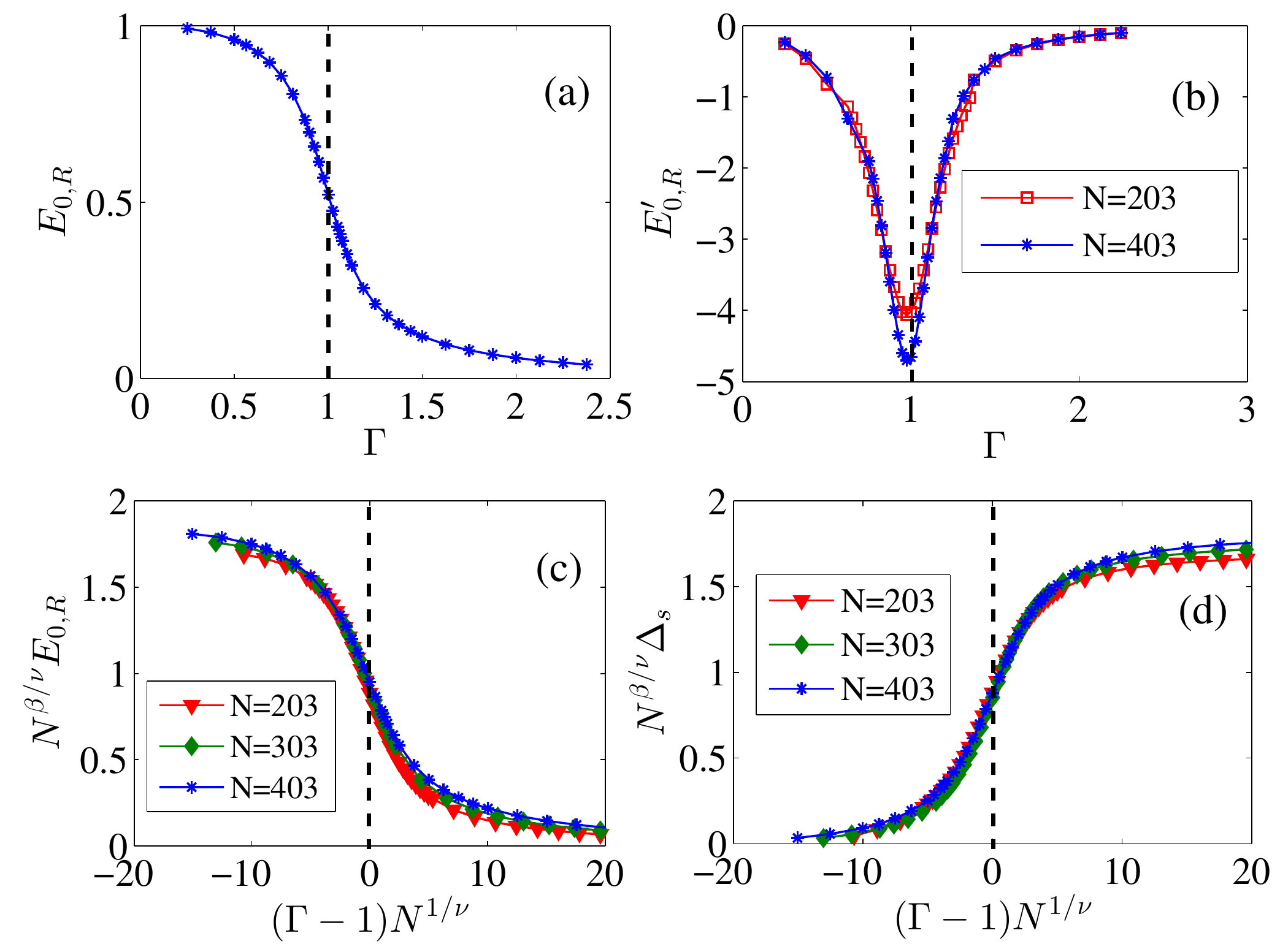}}
   \caption{{
     \protect\footnotesize
       (Color online) (a) Negativity between the impurity and the right channel (i.e. $E_{0,R}$) versus $\Gamma$ for $N=403$ and $J'=0.4$. (b) Derivative of $E_{0,R}$ with respect to $\Gamma$ for different system sizes. (c) Finite size scaling of $E_{0,R}$. (d) Finite-size scaling of the Schmidt gap.}}
\label{NegOrderPara}
\end{figure}

{\em Negativity as an order parameter.-} Several entanglement measures have
been used to detect quantum phase transitions
\cite{amico2008,de2012entanglement,bayat2014order,PhilippeWalter,KLH,StephanNicolas,FrancisKaryn}. Here, we propose the negativity \cite{Kim-Negativity-2000,vidal2002}  as an order parameter for the IQPT, with $\Gamma$ as control parameter. For any bipartite density matrix $\rho_{AB}$ the negativity, as an entanglement measure, is defined as $E_{A,B}=-1+ \sum_k |\eta_k|$, where $\eta_k$'s are the
eigenvalues of the matrix $\rho_{AB}^{T_A}$, where $T_A$ stands for partial transposition with respect to subsystem $A$ \cite{SupMat}.
In this section, and through the remainder
of the paper we use $H_{OBC}$, with $N_L=N_R$ {\it odd}.
In Fig.~\ref{NegOrderPara}(a) we plot the negativity between impurity and
right channel, $E_{0,R}$, versus  $\Gamma$. It is expected that the ground
state is overscreened only at $\Gamma=1$ where the impurity is entangled with
both channels.  For any $\Gamma \neq 1$ in the thermodynamic limit, the
impurity is screened only by the channel with the strongest coupling to the impurity, resulting in a fully screened 1CK
phase.  Indeed, the behavior of the negativity is consistent; $E_{0,R}$ goes
from $1$ to $0$ around the critical point. The thermodynamic behavior can be
explored by studying the derivative of the negativity with respect to $\Gamma$,
namely $E'_{0,R}$, shown in Fig.~\ref{NegOrderPara}(b).
As the figure shows, the derivative dips at the critical point with
the dip sharpening as $N$ increases. This
suggests that, as $N\to\infty$, $E'_{0,R}$
diverges at the critical point, implying that the 2CK
ground state is destroyed and 1CK physics is emerging.

The interpretation of the negativity as an order parameter can be justified by
a finite-size scaling analysis \cite{barber}. An order parameter scales as
$|\Gamma-1|^\beta$ in the vicinity of the critical point and the correlation
length as $|\Gamma-1|^{-\nu}$, where $\beta$ and $\nu$ are critical exponents.
The role of a correlation length is here taken by the critical crossover scale
$\xi^{\ast}$ at which the renormalization-group  flow of the channel-asymmetric
model crosses over from the unstable overscreened fixed point to the fully
screened Kondo fixed point
\cite{affleck1991critical,mitchell2011real}. Finite-size scaling \cite{barber} implies that
\begin{equation}
E_{0,R}=N^{-\beta/\nu}F\left(\left|\Gamma-1\right| N^{1/\nu}\right),
\end{equation}
with $F$ a scaling function. In Fig.~\ref{NegOrderPara}(c), we plot $N^{\beta/\nu} E_{0,R}$ as a function of $\left(\Gamma-1\right) N^{1/\nu}$.  When $\nu\!= \!2\pm0.05$ and $\beta\! = \!0.2\!\pm\!0.02$, curves for different $N$ collapse to a single curve. The value of $\nu\!\approx\!2$ matches  CFT \cite{affleck1991critical} and  bosonization results \cite{fabrizio1995}, verifying that the negativity behaves as an order parameter. Here $\nu \!=\!1/d$, with $d$ the scaling dimension of the relevant operator that appears in the Hamiltonian when parity symmetry is broken.

{\em Schmidt gap.-} Another key quantity, related to the entanglement spectrum,
is the Schmidt gap $\Delta_S$.  Given a bipartitioning of the system, it is
defined by $\Delta_S=\lambda_1-\lambda_2$, where $\lambda_1 \ge \lambda_2$ are
the two largest eigenvalues of the reduced density matrix of any of the two
subsystems. It was recently shown that the Schmidt gap can serve as an order
parameter across quantum phase transitions \cite{de2012entanglement,bayat2014order}. For the 2CK model close to $\Gamma =1$, and
choosing a bipartition as shown in Fig.~\ref{SCKM2CK}(c) for two complementary
left and right blocks, the Schmidt gap is found to obey finite-size scaling
with the same critical exponents as the negativity.  Fig.~\ref{NegOrderPara}(d)
shows the Schmidt gap data collapse for three different system sizes, confirming it as an alternative order parameter to the negativity in the 2CK model.
\begin{figure}[t]
\centering{
\includegraphics[width=0.52\textwidth]{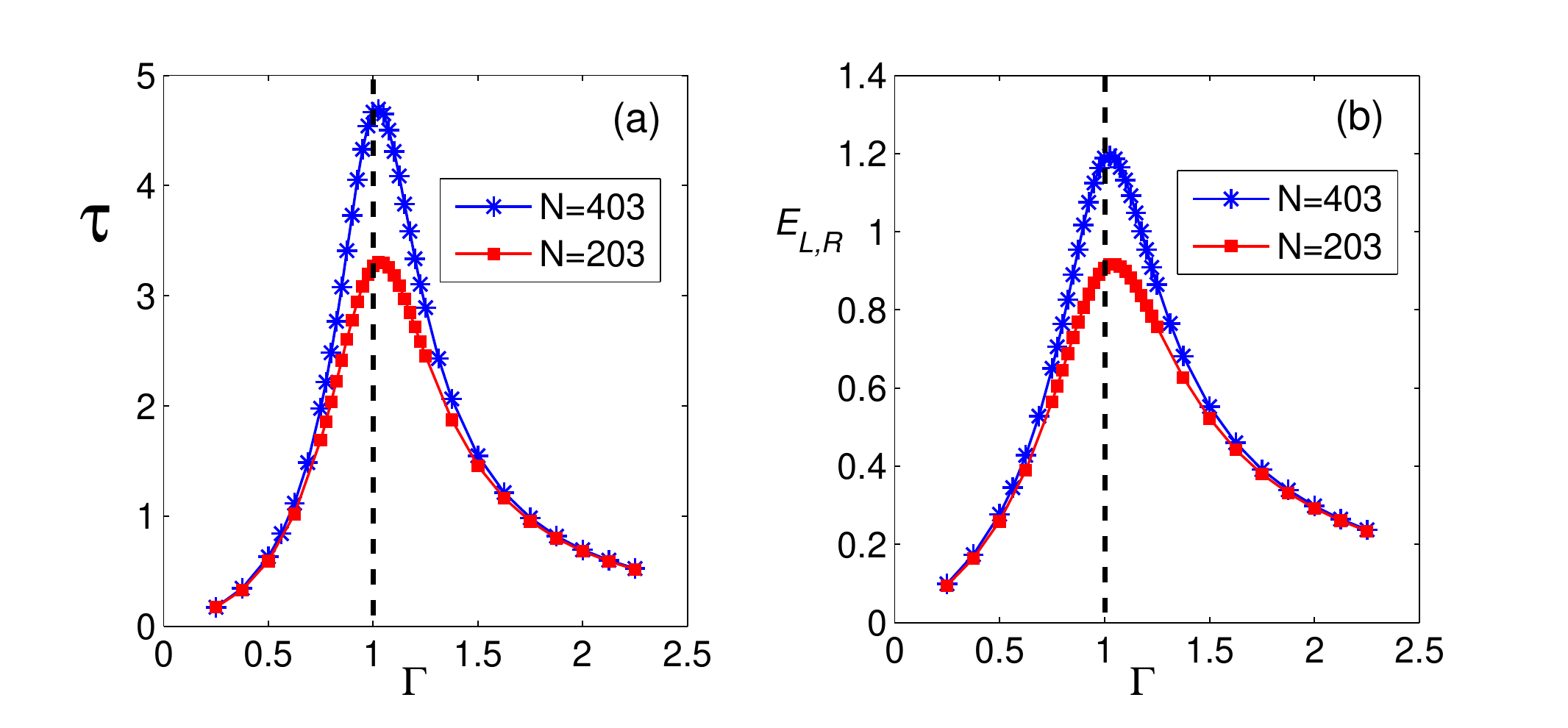}}
   \caption{{
     \protect\footnotesize
       (Color online) (a) The tripartite entanglement indicator $\tau$ versus $\Gamma$. (b) Entanglement between the two channels versus $\Gamma$. In both panels $J'=0.4$. }}
\label{tangle_fig}
\end{figure}

{\em Tripartite entanglement.-} Changing from 1CK to 2CK physics changes the entanglement structure fundamentally. Inspired by tangle~\cite{tangle-wootters} and its generalization for negativity~\cite{tangle-Fan},  as tripartite entanglement measures for qubits, we introduce a tripartite entanglement indicator as
\begin{equation}\label{tangle}
\tau=(\pi_0+\pi_L+\pi_R)/3
\end{equation}
in which
\begin{equation}\label{tangle2} \nonumber
\pi_0= E_{0,LR}^2-E_{0,L}^2-E_{0,R}^2, \text{       }
\pi_m= E_{m,0\overline{m}}^2-E_{m,0}^2-E_{m,\overline{m}}^2,
\end{equation}
where $m=L,R$ and $\overline{m}=R,L$ represent opposite channels, $E_{0,LR}=1$
is the negativity of the impurity with the rest of the system,
$E_{0,m}=E_{m,0}$ is the negativity between the impurity and channel $m$,
and $E_{m,\overline{m}}$ is the negativity between the two channels.
For systems with {\it odd} length leads each channel effectively behaves like a spin-1/2 system and our tripartite entanglement indicator $\tau$ becomes a natural generalization of the tangle defined for three qubits \cite{tangle-Fan}. In Fig.~\ref{tangle_fig}(a) we plot $\tau$ versus $\Gamma$ for systems with odd length leads.
$\tau$ clearly peaks at the critical point with the peak becoming more
pronounced with increasing length, suggesting its divergence with $N$.
The emergence of tripartite entanglement is therefore related to the overscreening
at the critical point where the two channels become highly entangled.
In Fig.~\ref{tangle_fig}(b), we plot the negativity between the two channels,
$E_{L,R}$, versus $\Gamma$.
As the figure
shows, $E_{L,R}$ is maximal at $\Gamma=1$, likely diverging with $N$.

{\em Conclusions.-}
Employing high-precision DMRG computations, we have studied the ground state entanglement of the 2CK model, allowing us to uncover the fractional ground state degeneracy predicted by CFT. The existence of the characteristic length scale $\xi_{2CK}$ is established through a
scaling analysis of $S_{imp}$.  The IQPT appearing as a function of channel-asymmetry and its exponents is detected using both the negativity and the Schmidt gap as order parameters.  Furthermore, the tangle is used to show that tripartite entanglement emerges only in the vicinity of the critical point.


\emph{Acknowledgements.-} The authors would like to thank E. Eriksson and N. Laflorencie for valuable discussions and
acknowledge the use of the UCL Legion High Performance Computing Facility (Legion@UCL), and associated support services, in the completion of this work.
Part of the calculations was made possible by the facilities of the Shared Hierarchical
Academic Research Computing Network (SHARCNET:www.sharcnet.ca) and Compute/Calcul Canada and some of the calculations were performed using the ITensor library~\cite{itensor}.
BA is funded by King Saud University. AB and SB are supported by the EPSRC grant EP/K004077/1.
IA is supported by NSERC Discovery Grant 36318-2009 and by CIFAR. ESS is supported by NSERC Discovery Grant.  SB also acknowledges support of the ERC grant PACOMANEDIA and EPSRC grant EP/J007137/1.  HJ acknowledges support from the Swedish Research Council and STINT. PS thanks the Ministry of Science, Technology and Innovation of Brazil, MCTI and UFRN/MEC  for financial support and CNPq for granting a "Bolsa de Produtividade em Pesquisa". KLH also acknowledges the CIFAR in Canada and KITP for hospitality.

\newpage

\appendix
\setcounter{figure}{0}
\setcounter{equation}{0}

\renewcommand{\theequation}{S\arabic{equation}}
\renewcommand{\thefigure}{S\arabic{figure}}
\onecolumngrid
\begin{center}
\LARGE{Supplemental Material}
 \end{center}

\section{1.  Review of negativity as an entanglement measure}

In a bipartite system $AB$, quantification of entanglement between the two subsystems has been intensively studied in the last decade.  When the density matrix of the overall system $\rho_{AB}$ is pure then the von Neumann entropy of the subsystem, namely $S(\rho_A)=-\text{Tr}\{\rho_A\log\rho_A\}$ where $\rho_A$ is the reduced density matrix of the subsystem $A$, is a unique measure of entanglement and all other measures are monotonic functions of the von Neumann entropy. At variance, when the overall state $\rho_{AB}$ is mixed the von Neumann entropy fails to quantify the entanglement between the two subsystems.
In such cases one can use negativity as a pertinent measure of entanglement, defined as \cite{Kim-Negativity-2000,vidal2002}
\begin{equation}\label{Negativity_definition}
E=-1+ \sum_k |\eta_k|=2 \sum_{\eta_k<0} |\eta_k|,
\end{equation}
where the $\eta_k$'s are the eigenvalues of the matrix $\rho_{AB}^{T_A}$ (or $\rho_{AB}^{T_B}$) in which $T_A$ (or $T_B$) stands for partial transpose with respect to the subsystem $A$ (or $B$). If the density matrix $\rho_{AB}$ is separable, then both $\rho_{AB}^{T_A}$ and $\rho_{AB}^{T_B}$ remain positive and thus  $\eta_k\geq 0$ for all $k$'s which results in zero negativity in Eq.~(\ref{Negativity_definition}). In contrast, if the overall state $\rho_{AB}$ is entangled then some of the $\eta_k$'s become negative and thus the negativity in Eq.~(\ref{Negativity_definition}) becomes nonzero.

Negativity as a measure of entanglement, applicable for both pure and mixed states, is an entanglement monotone which means that it does not increase under local operations. Furthermore, negativity is a legitimate quantum mechanical observable in the sense that it is associated with a Hermitian operator as
\begin{equation}\label{Negativity_observable2}
\mathcal{O}=2\sum_{\eta_k < 0}\left( |\eta_k \rangle \langle\eta_k | \right)^{T_A},
\end{equation}
where $|\eta_k\rangle$'s are the eigenvectors of the matrix $\rho_{AB}^{T_A}$. So, one can easily show that $E= \text{Tr} \left( \rho_{AB} \mathcal{O} \right).$

\section{2. Spin-chain representation of the 2CK model}

The spin-chain representation of the spin sector of the two-channel Kondo  model was first introduced in Ref. \cite{eggert1992magnetic,affleck1994}. To further justify this representation, we here provide a conformal field theory analysis. \\

By folding the system in half, creating an open boundary at the impurity site $l=0$, the left and right parts of the spin chain come to define two ''channels" on the interval $[0,Na/2]$, where $a$ is the lattice spacing. Introducing slowly varying left/right-moving SU(2)$_1$ spin currents  $\boldsymbol{J}^{(m)}_{LM/RM}$, together with two SU(2)$_1$ Wess-Zumino-Witten matrix fields $\boldsymbol g^{m}$ \cite{affleckLesHouches}, the representation
\begin{equation}
{\boldsymbol{\sigma}}^{(m)}_{l} \rightarrow \boldsymbol{J}^{(m)}_{LM}(al)  + \boldsymbol{J}^{(m)}_{RM}(al)  + (-1)^l \hbox{Tr}(\boldsymbol{g}^{(m)}(al) \boldsymbol{\sigma}),
\end{equation}
can then be used in the continuum limit to map the folded $\Gamma=1$ system onto the spin sector of the two-channel Kondo  model \cite{eggert1992magnetic,affleck1994}. Since the two channels (a.k.a. spin chains) both couple to the impurity, the expected chiral SU(2)$_1 \otimes$ SU(2)$_1$ symmetry of the critical theory in absence of the impurity gets replaced by SU(2)$_2 \otimes \mbox{Z}_2$, where the Ising symmetry group $\mbox{Z}_2$ encodes the presence of the two channels. This conformal embedding is different from the one used for the original two-channel Kondo  model \cite{affleck1991critical}, and reflects the fact that the two underlying bulk theories are different: In the two-channel Kondo  model  one is dealing with non-interacting electrons with two orbital channels, whereas in its spin-chain representation the impurity host is that of a one-dimensional half-filled Hubbard model (with charge gapped out) and with only a single orbital. The Ising $\mbox{Z}_2$ sector comes into play only off the two-channel Kondo  critical point, and then contributes a leading scaling operator of the same dimension as that from the "flavor" sector in the conformal embedding used for the two-channel Kondo  model \cite{affleck1991critical}. This testifies to the consistency of the spin chain representation. Note that the total central charge implied by the spin-chain representation on a half-line,
$c = c_{\text{SU(2)}} + c_{\text{Ising}} = 3/2 + 1/2 = 2$, becomes halved, $c=1$, when doubling the spatial degrees of freedom by unfolding the system back to the full interval $[-Na/2, Na/2]$.
The unfolded geometry is the one used in our DMRG computations, cf. Fig. 1. \\

\section{3. Use of periodic boundary conditions in Fig 2}
In most studies of the 2CK model there is no interaction between the two channels; however, for the results presented in Fig 2 the two channels are coupled through the periodic
boundary conditions. For the remainder of the results presented this is not the case. As long as $x\ll N/2$ this should not affect the results for large enough $N$ since in
the limit $N\to\infty$ results have to be independent of this boundary condition. Obviously, once $x$ approaches $N/2$ there is no reason to expect that we should correctly
represent 2CK physics. This can for instance be seen in the inset of Fig. 2(b) where $S_{\text{imp}}(x,N)$ does not approach $\ln(2)/2$ as $x$ approaches
$N/2$. It is then reasonable to ask why we use periodic boundary conditions since numerically it would be simpler to use open boundary conditions. The {\it major} reason for
this is that the correction to scaling terms appearing in Eq. (9), proportional to $\alpha$, are well understood~\cite{toappear}. These terms are more correctly seen as arising
from 1CK physics associated with the open boundary in the left panel of Fig.~2(b). In the presence of open boundary conditions several
other terms would appear dominating the 2CK physics we are trying to study and it is therefore a significant advantage to employ the periodic boundary conditions as was done
in  Fig. 2.
Furthermore, we note that $H_{PBC}$ as defined in Eq. (2) approach a completely uniform chain in the limit $J'\to 1$ which is another advantage when extracting $S_{\text{imp}}$.
For the results presented in Figs. 3,4 the above advantages of using $H_{PBC}$ are not pertinent and we have therefore used the more standard $H_{OBC}$.

\section{4. Limiting Behavior of $S_{imp}$ as $J'\to 1$}
As discussed in the main text
$S_{\text{imp}}(J'\to 1)=\ln(2)-\ln(g_{UV}/g_{IR})$ using Eq.~(4) and (7).
This relation takes the view point og 2CK physics. It is of course also possible
to remain with a purely spin-chain view point in which case it follows from
Eq.~(4) and (6) that
$S_{\text{imp}}(J'\sim 1,x,N)\sim -2\ln g=\ln( 2)/2$ for $1\ll x\ll N$.
These two results just re-express the basic relation
\begin{equation}
g_{UV}/g_{IR}=2g^2,
\end{equation}
discussed in detail in the main text.

\end{document}